\documentstyle[aps,bezier,epsfig,amsmath,amssymb]{revtex}         
\newcommand{\upd}{{\mathrm d}}
\renewcommand{\epsilon}{\varepsilon}

\begin{document}

\draft
\twocolumn[\hsize\textwidth\columnwidth\hsize\csname@twocolumnfalse\endcsname 

\title{ \bf Nonequilibrium Critical Dynamics of the 2D XY model}

\author{ Ludovic Berthier$^{1,2}$,
Peter C. W. Holdsworth$^{2}$,
and Mauro Sellitto$^{2}$}  

\address{$^1$D\'epartement de Physique des Mat\'eriaux, 
Universit\'e C.
Bernard and CNRS, F-69622 Villeurbanne, France} 

\address{$^{2}$Laboratoire de Physique, \'Ecole Normale Sup\'eieure de Lyon 
and CNRS, F-69007 Lyon, France }

\date{\today}

\maketitle

\begin{abstract}
The nonequilibrium critical  dynamics of the 2D XY model is investigated 
numerically through Monte Carlo simulations and analytically in the 
spin-wave approximation.
We focus in particular on the behaviour of the two-time response
and correlation functions and show that the ageing dynamics 
depends on the initial conditions.
The presence of critical fluctuations leads to non-trivial 
violations of the fluctuation-dissipation theorem apparently reminiscent 
of the three dimensional Edwards-Anderson spin glass model.
We compute for this reason the finite-size overlap probability distribution 
function and find that it is related to the finite-time 
fluctuation-dissipation ratio obtained in the out of equilibrium dynamics, 
provided that the temperature is not very low.
\end{abstract} 

\pacs{PACS numbers: 05.70.Ln, 75.40.Gb, 75.40.Mg
\hspace*{4.2cm} LPENSL-TH-15/2000} 
\pacs{
{\it Il y a du miracle dans toute co\"{\i}ncidence}, {\sc M.Yourcenar}.}  

\twocolumn\vskip.5pc]\narrowtext

\section{Introduction}

One of the most interesting problems in the study of glassy
systems~\cite{review_ageing} is the characterization of the
nonequilibrium dynamics in terms of equilibrium phase
space structure.
In nonequilibrium dynamics temporal correlations do not
show time translational invariance. 
In the so-called ageing regime
the correlation function $C(t,t_w)$
of any observable is a two-time function
depending on both a waiting time $t_w$ and
$t$, the total time. $C(t,t_w)$ and its conjugate linear response $R(t,t_w)$
are related in a suitable long-time limit by a generalized
form of the fluctuation-dissipation-theorem (FDT)~\cite{CuKu_prl,cuku,frme},
\begin{equation}
R(t,t_w) = \frac{x(C(t,t_w))}{k_{\rm B} T}
\frac{\partial C(t,t_w)}{\partial t_w} \,.
\label{xfdt}
\end{equation}
The factor $x(C(t,t_w))$, the fluctuation-dissipation ratio (FDR),
is equal to 1 at equilibrium and its departure from this value
characterizes the ageing dynamics of a system out of equilibrium.
A simple kind of ageing, with $x=0$, occurs
for example in an Ising ferromagnet quenched from a
random configuration into the ordered phase below the critical 
temperature~\cite{BBK}.
Generally, glassy systems are characterized by more complex
behaviour, with  the FDR  a non-trivial
function of $t$ and $t_w$~\cite{review_ageing}.                      

In Ref.~\cite{silvio}, Franz {\it et al.} have shown that in a certain
class of systems the FDR~\cite{CuKu_prl}
is closely related to the Parisi function
which measures the distribution of overlaps between pairs of pure equilibrium
states~\cite{Pa}.
The FDR is
linked to the Parisi function $P(q)$ through the
relation~\cite{silvio}
\begin{equation}
x(C) = \int_0^C \upd q P(q).
\label{theorem}
\end{equation}
The relevance of such a kind of `ergodic' result is obvious as it allows
one to predict the long time value of physical observables in a
non-stationary dynamical regime, in terms of static averages computed at
equilibrium, i.e. using the Gibbs measure.
From this equation, a complex form for $x(C)$ is related, through $P(q)$, to
a system with many pure equilibrium states.
The result $x=0$ for ferromagnets quenched below 
the critical point~\cite{BBK} is also
consistent with it, as in this case $P(q) = \delta(q-m^2)$.
Further, it has been tested numerically with success in finite-dimensional
spin glasses~\cite{fdt_ea2}.
However, many interesting situations also arise when the technical
hypothesis under which the result of Ref.~\cite{silvio} holds are not
strictly met. 
This occurs for instance when the
asymptotic value of the free-energy is different from that at
equilibrium~\cite{CuKu_prl,Zar,Struik}.
In these examples, violations of the fluctuation-dissipation
theorem reminiscent
of glassy systems may indeed appear even in systems with trivial Gibbs
measures, at least on finite but large time-scales~\cite{Se}.
In this case, one may wonder whether an appropriate generalization
of (\ref{theorem}) can be envisaged.

It has been recently suggested that non-trivial FDT violations
appear even in non-disordered systems such as a ferromagnet,
provided they are at criticality~\cite{golu,marco}.
This proposition would be particularly interesting if it led to
simple models with behaviour analogous to that of glassy systems, but
for which we have a detailed knowledge of the microscopic excitations.
In this paper we investigate this proposition by studying
the nonequilibrium critical dynamics of
the 2D XY model and its possible relation with the static 
properties.     

At the critical point $T=T_c$,
the dynamics of a ferromagnet undergoes so-called  `critical
slowing down'~\cite{HoHa,Ma}.
This is a consequence of the fact that at $T_c$
the equilibrium correlation length $\xi_{\rm eq}(T_c)$,
and hence  the largest relaxation time, diverges~\cite{HoHa,Ma}.
In particular, when a ferromagnet is quenched from a highly
disordered state to the critical point $T=T_c$~\cite{janssen},
the correlation length increases with the time $t$ elapsed after the
quench as $\xi(t) \sim t^{1/z}$, which defines the dynamical exponent $z$.
At time $t$, only the critical fluctuations with wavelength
smaller than $\xi(t)$ have equilibrated.
The magnetization over a length scale $\xi(t)$ is given by
$m_{\xi}(t) \sim \xi^{-\beta/\nu} \sim t^{-\beta/\nu z}$ and vanishes
at long time.
In the thermodynamic limit, the system does not equilibrate
since the equilibrium correlation length is infinite
at the critical temperature, $\xi_{\rm eq}(T_c)=\infty$.
This is rather different from the situation in which a
ferromagnet is quenched  {\it below} the critical temperature
to $T<T_c$~\cite{alan1}.
Then, well-defined domains with finite and opposite magnetization
$\pm m_{\rm eq}(T)$ appear and coarsen.
At long time $t$, the system enters a scaling regime where the typical
size $L(t)$ of a domain grows as $L(t) \sim  t^{1/z}$, where $z$ is the
dynamic critical exponent defined above~\cite{alan1}.
This scaling holds in the regime where $\xi_{\rm eq}(T) \ll L(t) \ll L$.
Here, although equilibrium is never reached in the thermodynamic limit,
$L \to \infty$, the loss of equilibrium is restricted to the domain walls
whose density becomes zero at long time. 
The thermodynamics is given by the bulk, which is in 
equilibrium~\cite{BBK}. 

The finite-size equilibrium fluctuations of the
magnetization $m$ of a critical system extend
over all the range of length scales
between the lattice spacing $a$ and the system size~$L$.
The probability distribution function of $m$, denoted by $Q(m,L)$, scales
then as $Q(m,L) \sim L^{\beta/\nu} {\cal Q}(mL^{\beta/\nu})$, where
$\beta$ and $\nu$ are the usual critical exponents associated with the
magnetization and the equilibrium correlation length,
respectively~\cite{Binder}.
At the critical point the scaling function ${\cal Q}$ is non-Gaussian
and markedly asymmetric, with large fluctuations below the
mean~\cite{Binder}.
Hence, for finite-size critical systems a non-trivial Parisi function,
which for pure ferromagnets is directly related to the magnetization
distribution, is expected to occur.
       
In relation (\ref{theorem}), it is understood that the left hand side
is computed in the limit of large waiting times, after the thermodynamic
limit is taken.
The right hand side is computed in an infinite system at equilibrium.
In this paper, we make the
conjecture that FDT violations on a finite-time scale $t_w$
are governed by the Parisi function $P(q,L)$ of a system of
finite-size $L$ such that $L=\xi(t_w)$.
This amounts to a finite-time, finite-size generalization
of the relation (\ref{theorem}), namely
\begin{equation}
x(C(t,t_w)) = \int_0^{C(t,t_w)} \upd q \, P(q,\xi(t_w)).
\label{conjecture}
\end{equation}
The right hand side is now computed by equilibrating a finite
system.

To test this proposition,
we focus on the static and dynamic properties of
the 2D XY model below the Kosterlitz-Thouless transition, where
the system is characterized by a line of critical points~\cite{koth}.
In this respect the 2D XY model is particularly interesting, since
the static distribution of the order parameter fluctuations and
the non-equilibrium dynamics can be analytically obtained in the
spin-wave approximation.
It has also the advantage that it is critical over all the
low-temperature phase and therefore does not require a temperature
fine-tuning in numerical simulations.
Moreover, a similarity between the 2D XY model
and the 3D Edwards-Anderson model has often been
noticed~\cite{iniguez}.
Actually, the apparent analogy might not be completely accidental:
it has been proposed that extremal statistics are relevant 
in disordered systems and turbulence~\cite{BoMe},
while $Q(m,L)$ in the 2D XY model is closely related
to a generalized Gumbel distribution~\cite{BHP_prl,peter} which also 
describes the power fluctuations in a confined-turbulence
experiment~\cite{BHP}.

The paper is organized as follows.
In the next section, we study analytically within
the spin-wave approximation
the behaviour of two-time correlation and response functions as well as the
FDT violations starting from a completely ordered state.
In Section III, the same quantities are investigated by Monte Carlo
simulations for both cases of random and ordered initial conditions.
This allows one to go beyond the spin-wave approximation and see how the vortex
contribution changes the dynamic scaling of two-time functions.
In Section IV, we compute the static fluctuations of the spin glass order
parameter in a finite size system and
test relation (\ref{conjecture}).
Our results are finally discussed  in Section V.

\section{Ageing dynamics: Analytical Results}

The coarsening dynamics of the 2D XY model has been largely studied
both numerically and 
theoretically~\cite{alan2,alan3,rutenberg,stefano,other1,other2,other3,other4,cukupa}.
These studies have mainly focused on the coarsening process itself,
rather than on the ageing properties (two-time functions and FDT violations).
The only paper dealing with FDT violations in the 
XY model is Ref.~\cite{cukupa}, but the spin-spin 
autocorrelation function,
together with its conjugated response function have not been studied.
We shall fill the gap in this Section, making use of the formalism
and results of Refs.~\cite{alan3,cukupa}.

\subsection{Dynamics in the spin-wave approximation}

The relaxational dynamics of the XY model can be described by
a Langevin dynamics~\cite{alan1}
\begin{equation}
\frac{\partial \boldsymbol{\phi}(\boldsymbol{x},t)}{\partial t}
= - \frac{\delta F [\boldsymbol{\phi}]}{\delta 
\boldsymbol{\phi}(\boldsymbol{x},t)} 
+ \boldsymbol{\zeta}(\boldsymbol{x},t).
\end{equation}
In this expression, $\boldsymbol{\phi}(\boldsymbol{x},t)$ is a two-component
order parameter, and the free-energy functional 
$F[\boldsymbol{\phi}]$ is of the Ginzburg-Landau type.
In the spin-wave approximation  the free energy 
functional reads~\cite{alan3}
\begin{equation}
F [\theta]= \frac{\rho(T)}{2} \int \upd^2 \boldsymbol{x} [\boldsymbol{\nabla} 
\theta(\boldsymbol{x})]^2,
\label{hamiltonian}
\end{equation}
where $\rho(T)$ is the spin-wave stiffness and
the angular variable $\theta$ is related to the spin variable 
by $\boldsymbol{\phi} \equiv \exp(i \theta)$.
In this approximation the dynamics is described by the following Langevin 
equation
\begin{equation}
\frac{\partial \theta(\boldsymbol{x},t)}{\partial t}  
= - \frac{\delta F[\theta]}{\delta \theta(\boldsymbol{x},t)} +
\zeta(\boldsymbol{x},t).
\label{dyn}
\end{equation}
In this expression, $\zeta$ represents the thermal noise.
It is a random Gaussian variable with mean
$\langle \zeta (\boldsymbol{x},t) \rangle=0$
and variance $\langle \zeta (\boldsymbol{x},t) \zeta 
(\boldsymbol{x'},t')\rangle
= 4 \pi \eta(T) \rho(T) 
\delta(\boldsymbol{x}-\boldsymbol{x'}) \delta(t-t')$, where
$\eta(T)$ is the usual critical exponent at temperature $T$,
linked to the spin-wave stiffness by the relation $2 \pi \eta(T) \rho(T) 
=T$~\cite{koth}. 

To be complete,
initial conditions have to be associated with this first order 
differential equation.
Since the Hamiltonian (\ref{hamiltonian}) only describes 
the spin-wave excitations of the system, 
the dynamics (\ref{dyn}) cannot be a good representation of 
the coarsening process following a quench from a completely
disordered state.
In that case vortices which are not described by the
spin-wave approximation are present in the system. 
We therefore take as initial conditions a completely
ordered configuration,
\begin{equation}
\theta(\boldsymbol{x},0) = \theta_0 \,,  \qquad
\forall \boldsymbol{x}\,.
\label{initial}
\end{equation}
Similar initial conditions have been studied in Ref.~\cite{cukupa}, but
one could generally study quenches from any temperature
below the transition, as was done in Ref.~\cite{alan3}.

The dynamics is easily solved by Fourier transforming the 
dynamical equation. 
This gives 
\begin{equation}
\theta(\boldsymbol{k},t) = \int_0^t \upd t' e^{-k^2 \rho(T) (t-t')} 
\zeta(\boldsymbol{k},t'),
\label{thetak}
\end{equation}
where $\theta(\boldsymbol{k},t) \equiv \int \upd^2 \boldsymbol{x}
\theta(\boldsymbol{x},t)e^{-i \boldsymbol{k} \cdot \boldsymbol{x}}$.
The initial condition (\ref{initial}), which in Fourier space reads
$\theta (\boldsymbol{k},0)=0$, has been used.

The spin-spin autocorrelation function is defined by
\begin{equation}
C(t,t_w) \equiv  \frac{1}{V} \int \upd^2\boldsymbol{x} \langle
\boldsymbol{\phi}(\boldsymbol{x},t) \cdot \boldsymbol{\phi}
(\boldsymbol{x},t_w) \rangle.
\end{equation}
It can be expressed in terms of the angular variables as
\begin{equation}
C(t,t_w) =   \frac{1}{V} \int \upd^2\boldsymbol{x} \langle 
  \cos [ \theta(\boldsymbol{x},t) - \theta(\boldsymbol{x},t_w)
 ]  \rangle .
\end{equation}
This quantity is easily computed by using the fact that 
$\theta(\boldsymbol{k},t)$ is a Gaussian variable, Eq.~(\ref{thetak}), 
by developing the trigonometric functions and using Wick's 
theorem~\cite{archambault}.
The correlation reads finally 
\begin{equation}
C(t,t_w) =   \left[
\frac{a^4}{(a^2+2t)(a^2+2t_w)} \cdot \frac{(a^2+t+t_w)^2}{(a^2+t-t_w)^2}
\right]^{\eta(T)/4}.
\label{corr1}
\end{equation}
An ultra-violet cutoff $a$, simulating the lattice spacing has been introduced
to regularize divergent expressions. This
is done, for convenience, by adding a factor $\exp(-k^2 a^2)$ in 
each integration over the Fourier space.

The conjugate response function is given by
\begin{equation}
R(t,t_w) \equiv  \frac{1}{V} \int \upd^2\boldsymbol{x} 
\frac{\delta \langle \boldsymbol{\phi} (\boldsymbol{x},t) \rangle}{\delta 
\boldsymbol{h}
(\boldsymbol{x},t_w)} {\Bigg \vert}_{\boldsymbol{h}=\boldsymbol{0}},
\end{equation}
where the field $\boldsymbol{h}(\boldsymbol{x},t)$ is conjugated with
$\boldsymbol{\phi}(\boldsymbol{x},t)$.
It may also be computed in terms 
of the angular variables, and in a similar way one finds
\begin{equation}
\begin{aligned}
R(t,t_w) = & \frac{1}{4 \pi \rho(T)(a^2+t-t_w)^{1+\eta(T)/2}} \times \\
& \times \left[ \frac{a^4 (a^2+t+t_w)^2}{(a^2+2t)(a^2+2t_w)}
\right]^{\eta(T)/4}
\end{aligned}
\end{equation}
We compute a two-time
fluctuation-dissipation ratio $X(t,t_w)$ defined as 
\begin{equation}
X(t,t_w)=\frac{TR(t,t_w)}{\frac{\partial C(t,t_w)}{\partial t_w}}\,.
\end{equation}
It is given by
\begin{equation}
X(t,t_w) =\left[ 1-\frac{(a^2+t-t_w)
(t-t_w)}{(a^2+t+t_w)(a^2 + 2 t_w)} \right]^{-1} .
\end{equation}

\subsection{Scaling behaviour}

The dynamic functions $C(t,t_w)$, $R(t,t_w)$ and $X(t,t_w)$ 
may be rewritten
in the following scaling forms, for times $t-t_w \gg a^2$:
\begin{equation}
\begin{aligned}
C(t,t_w) =  & \frac{1}{(t-t_w)^{\eta(T)/2}}  
\left[ \frac{(1+\lambda)^2}{4\lambda}
\right]^{\eta(T)/4}, \\
R(t,t_w) =  & \frac{1}{4 \pi \rho(T) (t-t_w)^{1+\eta(T)/2} } 
 \left[ \frac{(1+\lambda)^2}{4\lambda}
\right]^{\eta(T)/4}, \\
X(t,t_w) = & \left[  1- \frac{(\lambda - 1)^2}{2(1+\lambda)} \right]^{-1} \,,
\label{antiscaling}
\end{aligned}
\end{equation}
where $\lambda \equiv t/t_w$.
More precisely, in the time regime $a^2 \ll t-t_w \ll t_w$, i.e. $\lambda
\sim 1$,
\begin{equation}
\begin{aligned}
C(t,t_w) \sim & \frac{1}{(t-t_w)^{\eta(T)/2}}, \\
R(t,t_w) \sim & \frac{1}{(t-t_w)^{1+\eta(T)/2}}, \\
X(t,t_w) \sim & 1.
\label{equil}
\end{aligned}
\end{equation}
This is the quasi-equilibrium regime.
The dynamic functions decay at long 
time separations $t-t_w \gg t_w$ ($\lambda \gg 1$)
as power laws
\begin{equation}
\begin{aligned}
C(t,t_w) \sim & \frac{1}{t^{\eta(T)/4}}, \\
R(t,t_w) \sim & \frac{1}{t^{1+\eta(T)/4}}, \\
X(t,t_w) = & \tilde{X} (\lambda) > 1.
\label{noequil}
\end{aligned}
\end{equation}
The crossover between the two regimes takes place at time
$t - t_w \sim t_w$.
Hence, the correlation functions at different waiting times do not superpose:
the system ages.

\begin{figure}
\begin{center}
\psfig{file=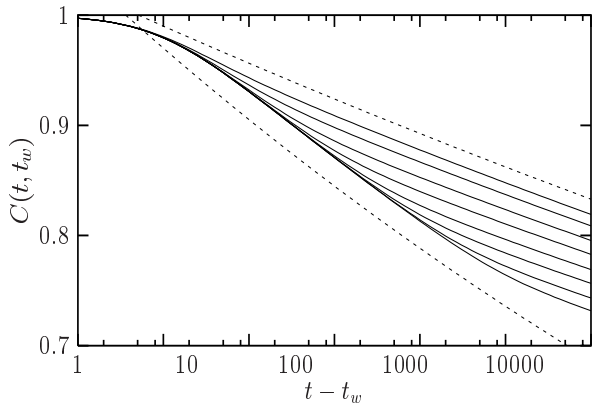,width=8.5cm,height=6.cm}
\caption{Analytical result (\ref{corr1}) for the correlation functions
for ordered initial conditions and different $t_w$. 
The dashed lines are power laws with exponents $-\eta/2$ 
(quasi-equilibrium part) and $-\eta/4$ (ageing part).
We have chosen $a=1$, $\eta/2=0.03$, and $t_w=1,\,3,\,10,\cdots,\,3000$,
from top to bottom.}
\label{CT03_3}
\end{center}
\end{figure}

\begin{figure}
\begin{center}
\psfig{file=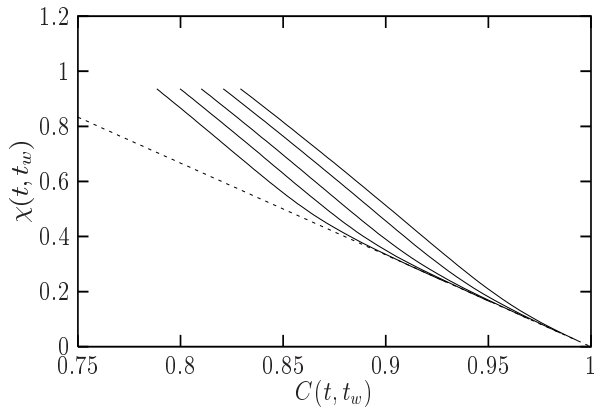,width=8.5cm,height=6.cm}
\caption{Analytical result for the parametric Susceptibility/Correlation
 plot, with 
$a=1$, $\eta/2=0.03$ and $t_w=1,\,3,\,10,\,30$ and 100, from top to bottom.
The dashed line is the equilibrium FDT between $\chi$ and $C$.}
\label{FDT}
\end{center}
\end{figure}

This scaling is in complete agreement with the results obtained by 
Godr\`eche and Luck~\cite{golu} for the spherical model at the 
critical point.
Its physical interpretation is straightforward.
At time $t_w$, fluctuations of wavelength smaller than
$\xi(t_w)$ have equilibrated, while others have not.
The quasi-equilibrium regime corresponds then
to equilibrium fluctuations of small wavelengths.
At equilibrium, the correlation decays as 
$C_{\rm eq}(t) \sim t^{-2\beta/\nu z}$.
For the 2D XY model~\cite{villain}, one has $\beta/\nu = \eta(T) /2$, $z=2$;
hence the 
exponent describing the stationary regime of the correlation function
is $- 2\beta/\nu z = -\eta(T)/2$, as found in Eq.~(\ref{equil}).

At the crossover time $t-t_w \sim t_w$, the autocorrelation
has the value $C(t_w+t_w,t_w) \sim t_w^{-\eta(T)/2}$.
It is interesting to compare this result with the case of 
ferromagnets quenched below $T_c$ where the 
crossover takes place for $C(t_w+t_w,t_w) \sim m_{\rm eq}(T)^2$.
In this case the correlation function displays a plateau at the 
value $q_{EA} \equiv m_{\rm eq}(T)^2$, where
$q_{EA}$ is the Edwards-Anderson parameter.
There is then a clear time separation
between the fast thermal fluctuations of very short wavelength
(recall that $\xi_{\rm eq}(T) \ll t^{1/z}$) and the slow
relaxation due to the domain wall motion.
For $T=T_c$, there is no such distinction but only 
a `pseudo-plateau' in the correlation, in the
sense that the crossover takes place at a waiting time dependent 
value of the correlation.
This reflects the fact that the order 
parameter vanishes in the thermodynamic limit.

The last regime, which takes place at a $t_w$-dependent time scale
is the ageing regime. 
It is characterized by the loss of time-translation invariance and
a breakdown of the FDT, $X(t,t_w) \neq 1$.
On this time scale, the decorrelation results from the growth
of the correlation length.
Note finally that although the response function is 
the product of a stationary and an ageing  part, the latter is a very 
slowly varying function of the variable $t/t_w$.
In a plot of $\chi(t,t_w)$ vs $t-t_w$,
there would be  apparently no ageing in the response function.

For comparison with the numerical results of the full  XY model, 
we show in Fig.~\ref{CT03_3} the analytical result (\ref{corr1}) 
for the correlation functions, while in Fig.~\ref{FDT} we build 
the usual plot to study the FDT violations~\cite{cuku}.
We have plotted the susceptibility $\chi(t,t_w) \equiv
\int_{t_w}^t \upd t' R(t,t')$ as a function of $C(t,t_w)$
parameterized by the time difference $t-t_w$, for different $t_w$.
With the generalized form of the FDT, Eq.~(\ref{xfdt}), the relation between
$\chi$ and $C$ becomes
\begin{equation}
\begin{aligned}
\chi(t,t_w) = & \frac{1}{T} \int_{t_w}^{t} \upd t' X(t,t') 
\frac{\partial C(t,t')}{\partial t'} \\
= & \frac{1}{T} \int_{C(t,t_w)}^1 \upd C' x(C'),
\label{plot}
\end{aligned}
\end{equation}
where the second line holds when the FDR becomes a function of $C$ only.
In this last case, the parametric plot reaches a master curve 
$\tilde{\chi}(C)$ independent of $t_w$, 
whose derivative $\upd \tilde{\chi}/\upd C$ is given
by $x(C)$.
At equilibrium, $x=1$ and $\tilde{\chi} (C)= (1-C)/T$.
When the FDR is not a single variable function, only
the first line of Eq.~(\ref{plot}) can be used and 
the FDR $X(t,t_w)$ is {\it not} the slope of the parametric plot.
It is clear from Eqs.~(\ref{antiscaling}) that this happens
in the 2D XY model, and a master curve $\tilde{\chi}$ is not reached.

Finally, we comment on the rather peculiar fact that 
in the time regime $\lambda \gg 1$ the FDR is 
larger than 1, which means from Eq.~(\ref{plot})
that the susceptibility is larger than its equilibrium value, 
as can be seen in Fig.~\ref{FDT}.
When the system is quenched, all degrees of freedom which were 
equilibrated at temperature $T=0$, fall 
out of equilibrium at time $t=0$.
As $t$ increases, `fast' degrees of freedom equilibrate, while
the `slow' ones are still out of equilibrium.
According to the  interpretation of the 
FDR as an effective temperature~\cite{cukupe}, 
$x>1$ means that the slow degrees of freedom 
are `quasi-equilibrated' at a temperature $T_{\text {eff}}\equiv T/x < T$.
Applied to a quench from a high temperature state, this reasoning 
implies a FDR smaller than one, and the recovery of the usual case $x<1$.
This argument is only qualitative, since the interpretation
of FDT violations in terms of an effective temperature relies on
the existence of well separated time scales, each one associated with
an effective temperature;
here instead there is no clear distinction between  `slow' and `fast' 
degrees of freedom, since there are fluctuations of every length scales, 
each having its own relaxation time.

\section{Ageing dynamics: Monte-Carlo Results}

In this section we present the results of Monte-Carlo simulations 
for the full XY model which allows to go beyond the spin-wave approximation.
We  first consider a quench from a completely ordered initial condition
to check the validity of the scalings obtained in the previous Section.
Then we consider a quench from a random initial condition in order 
to see how the vortex contribution changes the dynamical scaling
and the FDT violations.

\subsection{Model and details of the simulation}

The two-dimensional XY model is defined by the Hamiltonian
\begin{equation}
H= - \sum_{\langle i,j \rangle} \boldsymbol{\phi}_i \cdot
\boldsymbol{\phi}_j,
\end{equation}
where the sum runs over nearest neighbors of a square lattice of
linear size $L$, and  the spins 
$\{ \boldsymbol{\phi}_j, \, j=1,\, 2, \dots, \, L^2 \}$ are 
bidimensional vectors of unit length. 
Introducing angular variables through $\boldsymbol{\phi}_j \equiv
\exp ( i \theta_j)$, the Hamiltonian becomes
\begin{equation}
H = - \sum_{\langle i,j \rangle}  \cos(\theta_i - \theta_j).
\label{model}
\end{equation} 
The Kosterlitz-Thouless transition temperature is located at
$T_{KT} \simeq 0.89$~\cite{gupta}.
The model (\ref{model}) is simulated by the following Monte Carlo 
algorithm: the angular variable $\theta_i$ associated to a randomly
chosen site $i$ is updated to a new value $\theta_i'$ randomly
chosen in the interval $[- \pi,\pi ]$, with probability
$p=\min \left\{ 1,e^{-\beta \Delta E} \right\}$, where $\Delta E$ is the energy
variation between the two configurations.
We present results for a system of linear size  $L=512$.
We carefully checked that we are in a regime
with $\xi(t) \ll L$.

We shall be interested in the spin-spin autocorrelation function
$C(t,t_w)$ and in its conjugated integrated response 
function $\chi(t,t_w)$.
The former is easily computed from 
\begin{equation}
C(t,t_w) =  \frac{1}{L^2} \sum_{i=1}^{L^2} \langle \cos 
( \theta_i(t) - \theta_i(t_w)) \rangle,
\end{equation}
where the brackets denote an average over the realizations
of the thermal noise.
The computation of the latter is standard~\cite{BBK}. At time
$t_w$, a perturbation $\Delta H = - \sum_i \boldsymbol{h}_i \cdot
\boldsymbol{\phi}_i$ is added to the Hamiltonian.
The magnetic field is random: its two components are
independently drawn from a bimodal distribution $\pm h$.
In the linear response regime, the integrated response
function reads
\begin{equation}
\chi(t,t_w) = \frac{1}{h^2\,L^2} \sum_{i=1}^{L^2}
\overline{\langle \boldsymbol{h}_i \cdot \boldsymbol{\phi}_i \rangle},
\end{equation}
where the overline means an average over the realizations of the
random magnetic field.
We checked that the value $h=0.04$ was small enough for the
response to  be linear.

\subsection{Ordered initial conditions}

We investigate first the dynamic behaviour of the model (\ref{model})
starting from a completely ordered initial condition.
At $t=0$, the temperature is switched to $T < T_{KT}$.
The evolution of the system may be followed 
in a series of snapshots in Fig.~\ref{fieldb}.
At short times, the system is very ordered, and there is a single
color in the first snapshot.
As $t$ increases, fluctuations appear, and there is clearly a
growing length scale in the system.

\begin{figure}
\begin{center}
\begin{tabular}{ccc}
\psfig{file=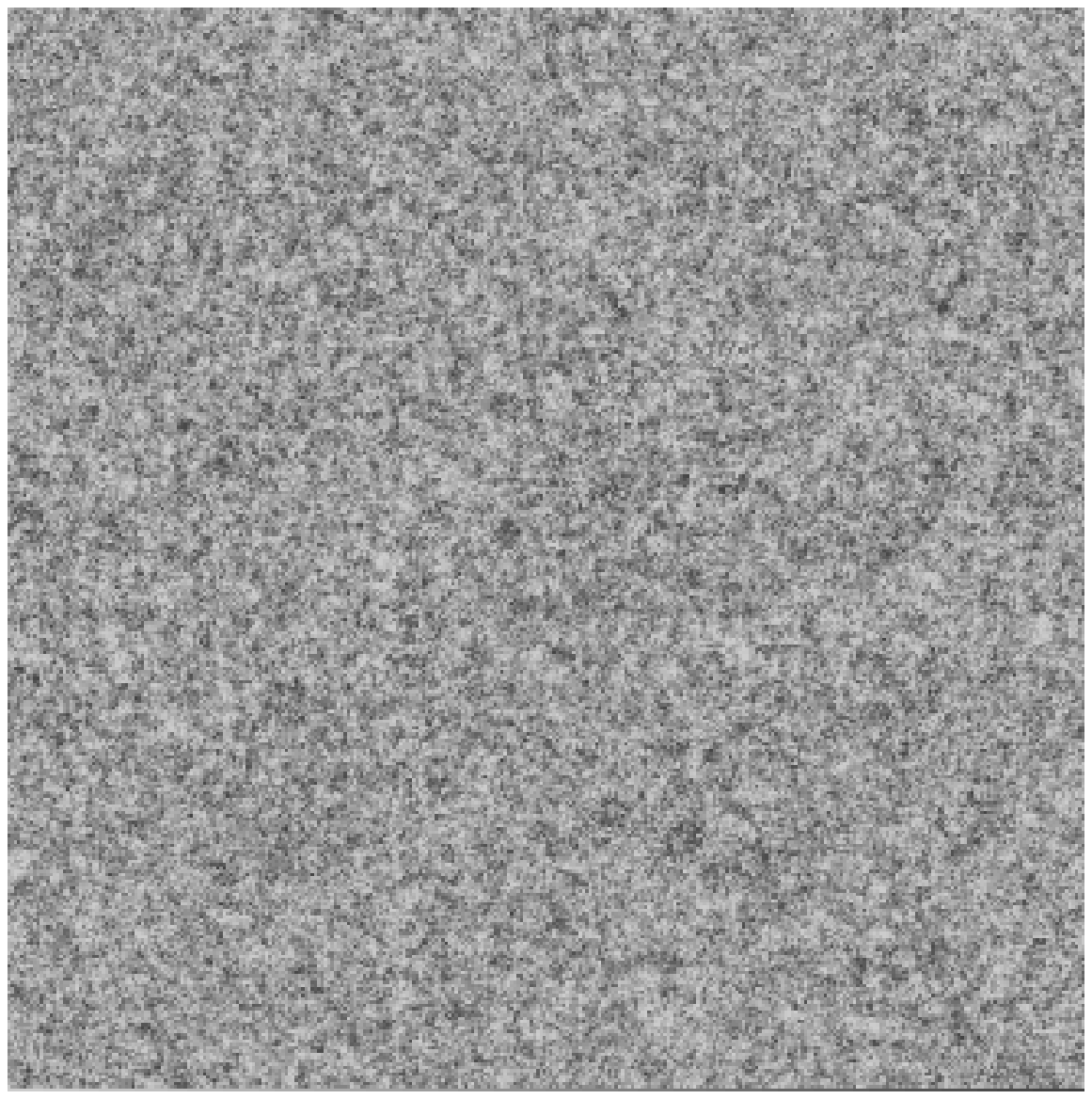,width=2.8cm,height=2.8cm} &
\psfig{file=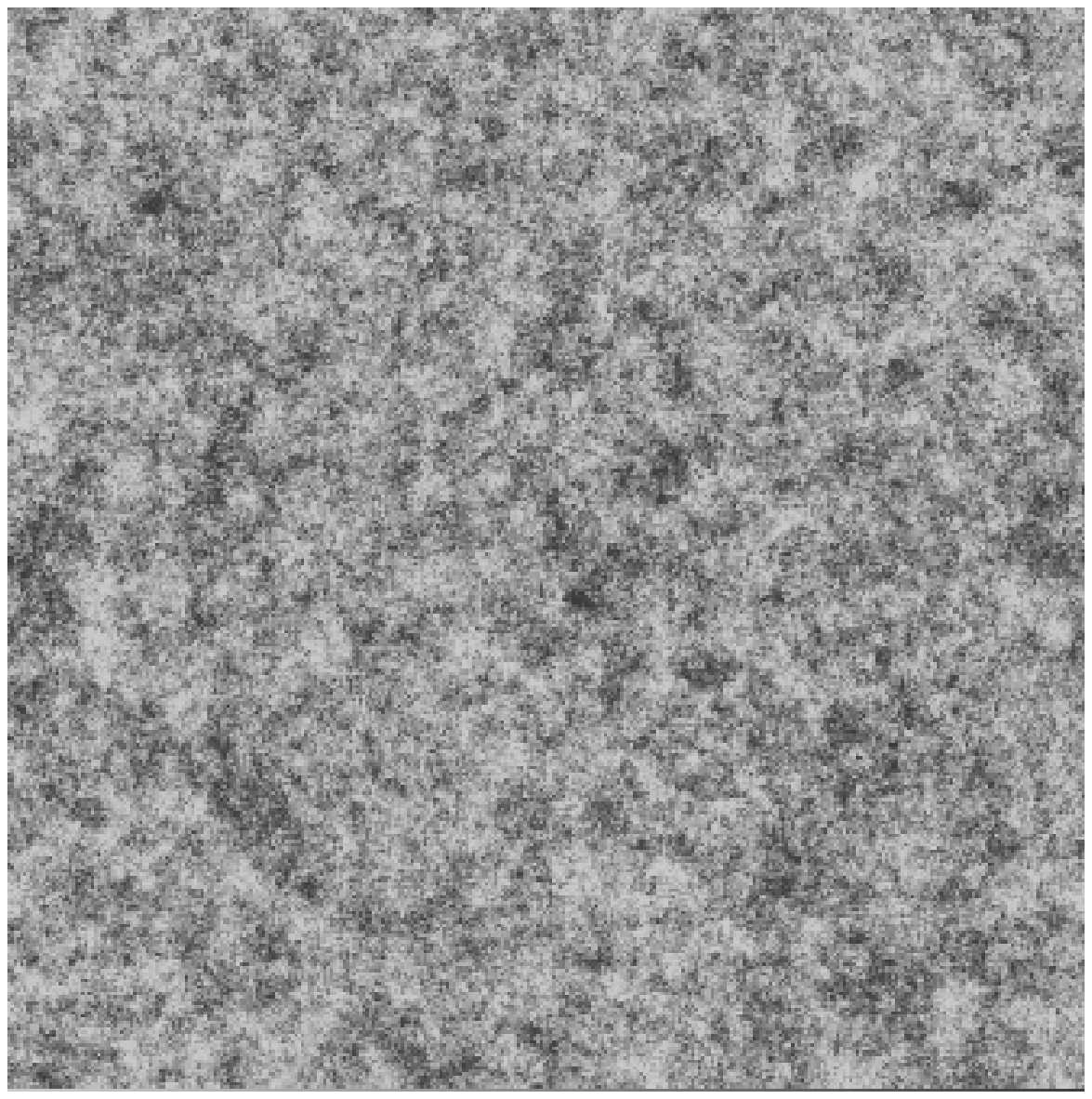,width=2.8cm,height=2.8cm} &
\psfig{file=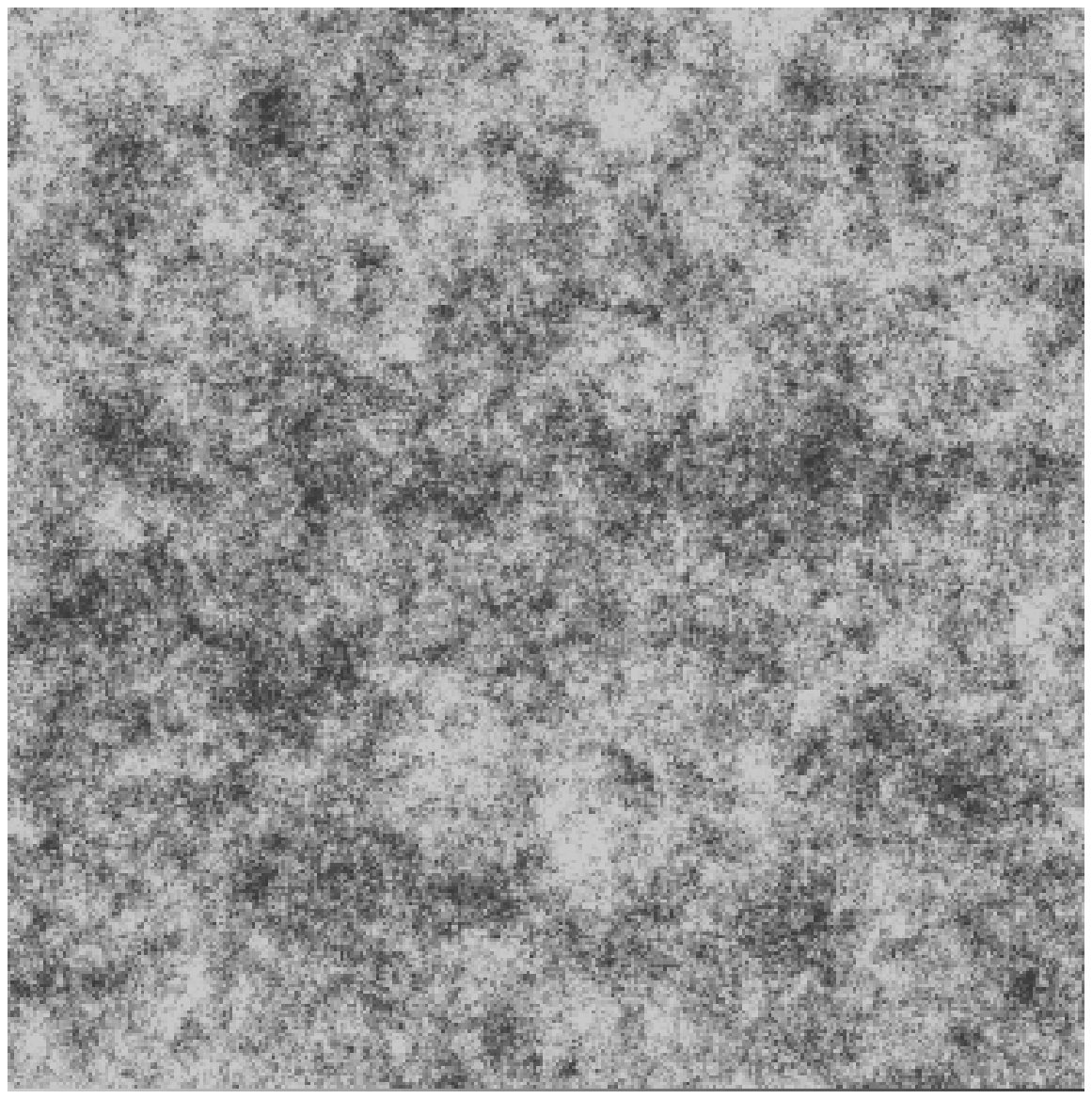,width=2.8cm,height=2.8cm} \\
\end{tabular}
\caption{Snapshot of the domain growth in the XY model at
$T=0.3$ for times $t_w = 10^2$, $10^3$, $10^4$ and ordered initial
conditions.
The system size is $512 \times 512$. The grey scale encodes the variable
$\cos(\theta(\boldsymbol{x},t))$.}
\label{fieldb}
\end{center}
\end{figure}

\begin{figure}
\begin{center}
\psfig{file=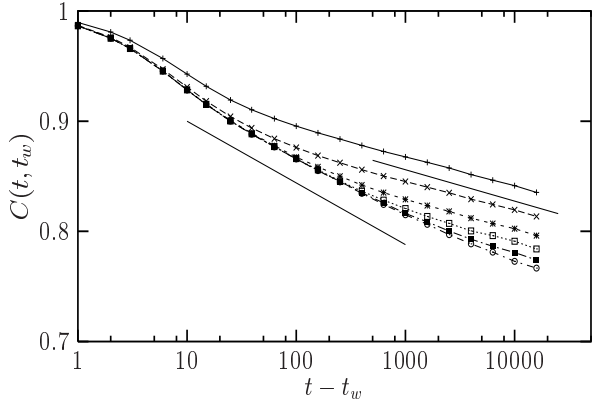,width=8.5cm,height=6.cm}
\caption{Autocorrelation function for ordered initial conditions,
for waiting times $t_w=10$, 30, 100, 3000, 10000 (from top to bottom).
The final temperature is $T=0.3$.
The full lines are power laws with exponent $-\eta/2=-0.03$ 
and $-\eta/4=-0.015$.}
\label{CT03_2}
\end{center}
\end{figure}

\begin{figure}
\begin{center}
\psfig{file=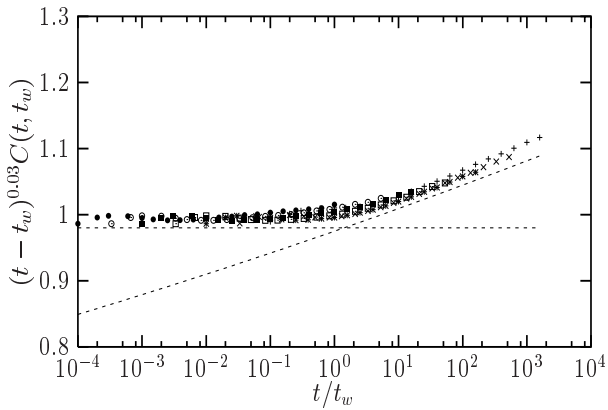,width=8.5cm,height=6.cm}
\caption{Rescaling of the 
autocorrelation functions
of Fig.~\ref{CT03_2}. The dashed lines are power laws with 
exponents 0 and 0.015.
The rescaling is good, except for the smallest waiting time.}
\label{SCT03_2}
\end{center}
\end{figure}

In Fig.~\ref{CT03_2}, the autocorrelation $C(t,t_w)$
is shown for the temperature $T=0.3$ and different
waiting times $t_w$.
The overall behaviour of $C(t,t_w)$ is in nice agreement with
the analytical results obtained in the spin-wave approximation, 
recall Fig.~\ref{CT03_3}.
As described in the previous section, there are two different
regimes, depending on the time scale.
For small time separation $t-t_w \ll t_w$, $C(t,t_w)$ exhibits
a stationary part, which is well represented by a power law
\begin{equation}
C(t,t_w) \sim \frac{1}{(t-t_w)^{0.03}},
\label{powerlaw}
\end{equation}
while at long time, the different curves do not superpose: the system
ages.
We test the scaling (\ref{corr1}) in Fig.~\ref{SCT03_2}, where
we plot $(t-t_w)^{0.03} C(t,t_w)$ as a function of the scaling variable
$t/t_w$.
We have actually obtained the exponent 0.03
as the value insuring the best collapse of these data.
This value of the exponent can be compared to the self-consistent
computation of Ref.~\cite{villain} which gives
$\eta(T=0.3) \simeq 0.026$. 
It is in good agreement with its
numerical estimation~\cite{gupta}.
At $T=0.7$, we find $\eta/2=0.085$, in agreement with the numerical 
estimation of Ref.~\cite{gupta}; the self-consistent
computation gives 0.069.

The parametric plot obtained numerically at $T=0.3$
for ordered initial conditions 
is represented in Fig.~\ref{FDT2}. It 
is also very similar
to the analytical results of Fig.~\ref{FDT}, confirming in particular
the unusual feature of a FDR larger than 1.  

The conclusion of the numerical study of ordered initial 
conditions is that the off-equilibrium dynamics
is satisfactorily described by the spin-wave approximation.
This is no longer true with random initial configurations 
on which we focus now.

\begin{figure}
\begin{center}
\psfig{file=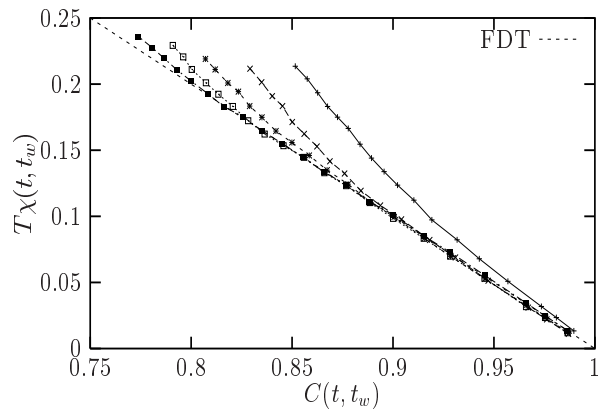,width=8.5cm,height=6.cm}
\caption{Parametric Susceptibility/Correlation 
plot for ordered initial conditions. 
The symbols are the same as in Fig.~\ref{CT03_2}.}
\label{FDT2}
\end{center}
\end{figure}

\subsection{Random initial conditions}

The non-equilibrium dynamics for random initial conditions 
turns out to be quite different.
However, our previous analytical results can be used to
understand the scaling behaviour of the dynamic functions
in that case too.

\begin{figure}
\begin{center}
\begin{tabular}{ccc}
\psfig{file=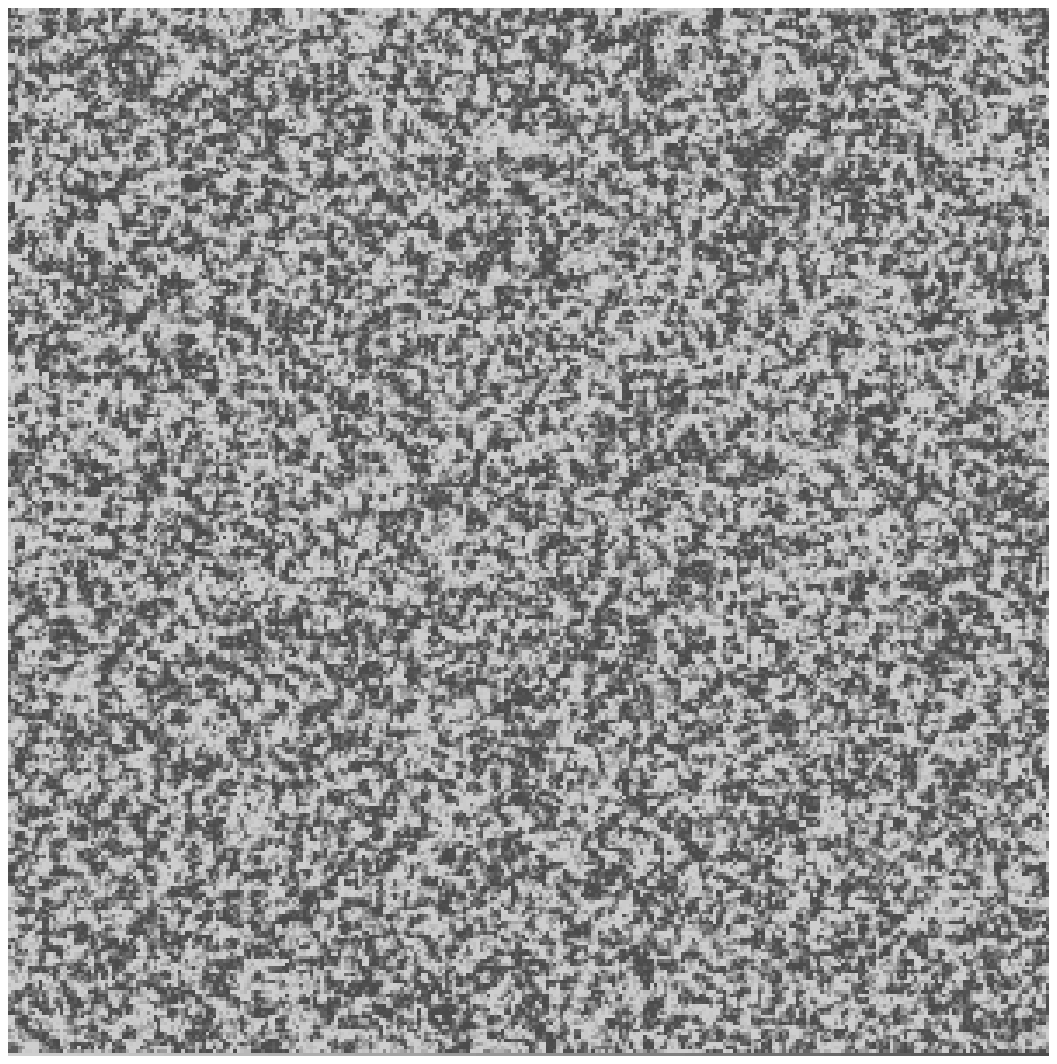,width=2.8cm,height=2.8cm} &
\psfig{file=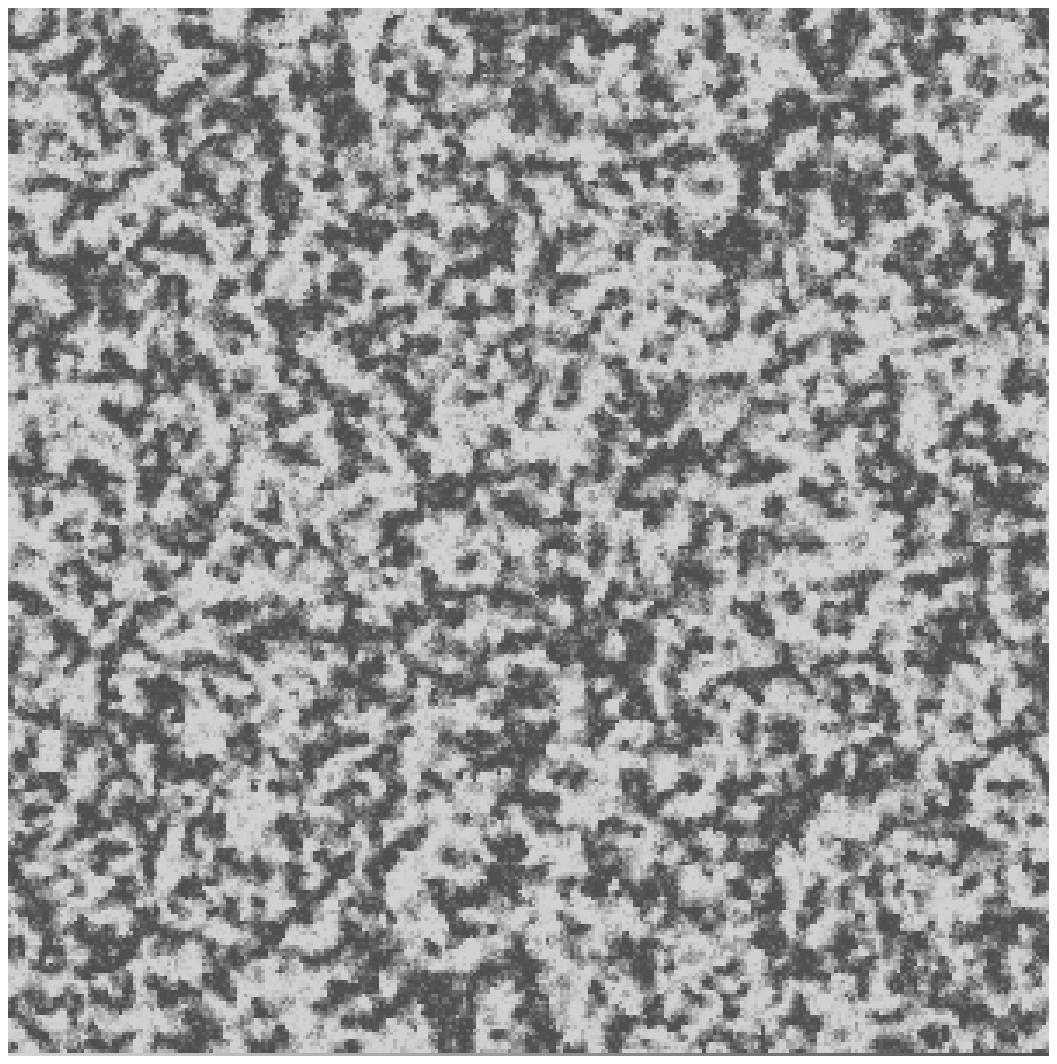,width=2.8cm,height=2.8cm} &
\psfig{file=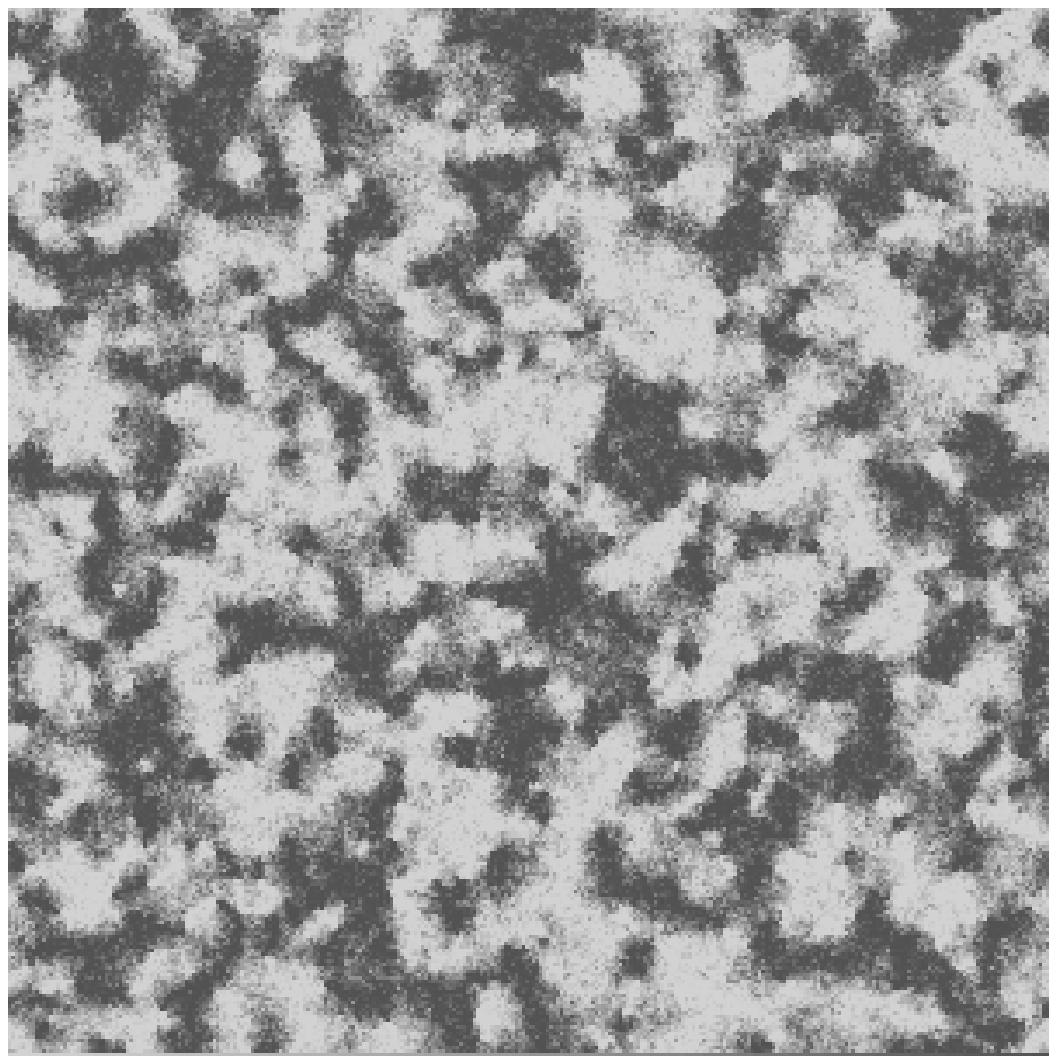,width=2.8cm,height=2.8cm} \\
\end{tabular}
\caption{Snapshot of the domain growth in the XY model at
$T=0.3$ for times $t_w = 10$, $10^2$, $10^3$ and random 
initial conditions.
The system size is $512 \times 512$. The grey scale encodes the variable
$\cos(\theta(\boldsymbol{x},t))$.}
\label{field}
\end{center}
\end{figure}

In Fig.~\ref{field} we present a series of snapshots for different
times. 
A simple look at Figs.~\ref{field} and \ref{fieldb}
makes the difference between the two situations very clear.
The snapshots of Fig.~\ref{field} are moreover different from the
`Schlieren patterns'  presented in $T=0$
simulations of the 2D XY 
model~\cite{rutenberg,stefano,other3} and
resulting from the presence of vortices.
In the present work, vortices are also present, but
they are blurred by thermal fluctuations, and do not clearly appear
in the snapshots, except at $T$ close to zero.

In Fig.~\ref{CT03}, we present the correlation function $C(t,t_w)$
at $T=0.3$ and different waiting times $t_w$.
The two distinct behaviours, namely a stationary part and an ageing
part, are still present.
The stationary part, corresponding to the equilibrium fluctuations,
remains unchanged and can again be represented by the power law
(\ref{powerlaw}).

\begin{figure}
\begin{center}
\psfig{file=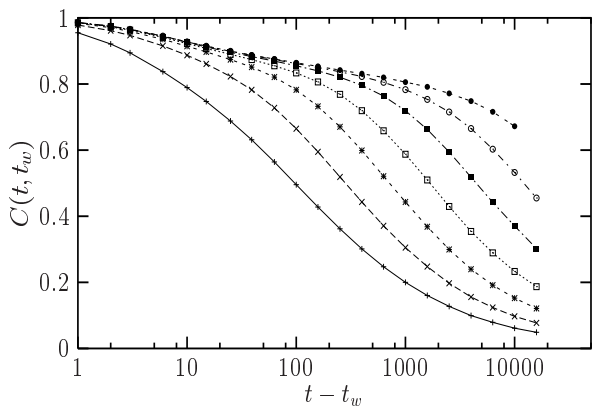,width=8.5cm,height=6.cm}
\caption{Correlation functions for random initial conditions at $T=0.3$ and
waiting times $t_w=10$, 30, 100, 300, 1000, 3000 and 10000 from
bottom to top.}
\label{CT03}
\end{center}
\end{figure} 

\begin{figure}
\begin{center}
\psfig{file=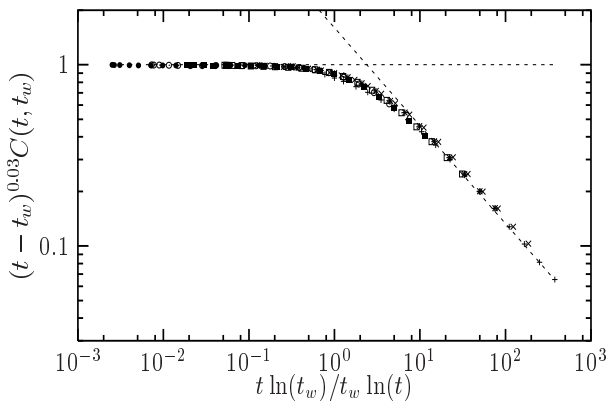,width=8.5cm,height=6.cm}
\caption{Rescaled correlation functions for random initial conditions
at $T=0.3$; the data are the  same as in Fig.~\ref{CT03}. 
The two dashed lines are power laws with exponents 0 and -0.54.}
\label{SCT03}
\end{center}
\end{figure} 

The ageing part is different because of the presence of vortices in 
the system.
By analogy with the previous case, 
we conjecture that $C(t,t_w)$ has the scaling form
\begin{equation}
C(t,t_w) = \frac{1}{(t-t_w)^{\eta(T)/2}} \Phi \left(
\frac{\xi(t)}{\xi(t_w)} \right),
\label{autocorr}
\end{equation}
where $\Phi$ is a scaling function and $\xi(t)$ is the growing
correlation length.
Previous numerical and analytical studies have shown
that an effect of the vortices is to change the growth law
for the correlation length from the power law $\xi(t) \sim t^{1/z}$
to the form $\xi(t) \sim \sqrt{t/ \ln t}$~\cite{alan2,rutenberg,other3}.
The scaling prediction for $C(t,t_w)$ is tested in Fig.~\ref{SCT03}
where we plot $(t-t_w)^{0.03}C(t,t_w)$ as a function of 
$\xi(t)/ \xi(t_w)$ including the logarithmic correction due
to the vortices.
The rescaling is very good. It is seen in particular that 
the long time decay is well described by a power law,
{\it i.e.} $\Phi(x) \sim x^{-\lambda}$ for large $x$.
At $T=0.3$, we find $\lambda=0.54$.
An asymptotic power law behaviour was also found in Ref.~\cite{golu} 
for the ferromagnetic spherical model.

It is also interesting to note that the $t/t_w$-scaling, 
usually called `simple ageing'~\cite{review_ageing} and  encountered 
for an ordered initial state, 
is changed to a `non-simple' ageing with the scaling 
$t\ln t_w /t\ln t $ when topological defects (here vortices) are present.
Although the two scalings are equivalent in the large waiting 
time limit, the latter is preasymptotically equivalent to a sub-ageing 
behaviour~\cite{review_vincent}.

\begin{figure}
\begin{center}
\psfig{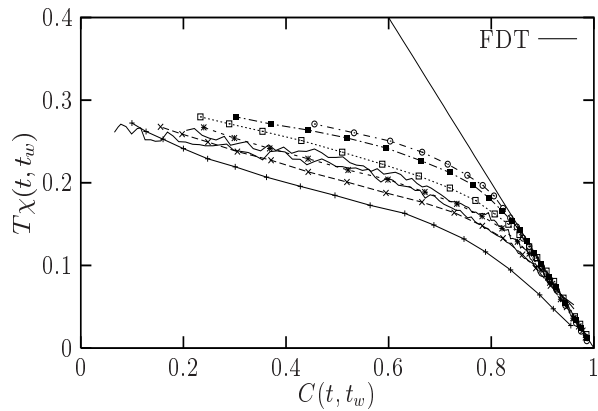} 
\caption{Parametric Susceptibility/correlation 
plot for $T=0.3$ and random initial conditions.
The symbols are the same as in Fig.~\ref{CT03}.
The two full lines are similar data obtained with the 3D Edwards-Anderson 
model (see text).}
\label{param}
\end{center}
\end{figure} 

We turn now to the FDT violations.
We present in Fig.~\ref{param} the parametric plot Susceptibility/Correlation.
There are still two distinct behaviours depending on the time scale.
At short time separation FDT holds, whereas at long time separation
a non-equilibrium regime is entered.
This plot is very different from the one obtained 
in Fig.~\ref{FDT} for ordered initial conditions in the sense that 
the susceptibility is smaller than the value obtained from
the FDT. In short, this means that $X(t,t_w) < 1$.
In the qualitative picture developed above in terms of 
an effective temperature, this means that the fluctuations of wavelength
larger than $\xi(t_w)$ are `quasi-equilibrated' at a temperature
higher than the thermal bath temperature, keeping then 
the memory of their $T=\infty$ initial state.

It is interesting to note the close similarity of this plot 
with those observed in simulations of the 3D Edwards-Anderson 
model~\cite{fdt_ea2,fdt_ea,BBK2}.
We have superposed as 
continuous lines the data obtained in Ref.~\cite{BBK2}.

\section{The link between statics and dynamics}

In this Section we compute the finite-size Parisi function
and compare it with the finite-time fluctuation-dissipation ratio
obtained in the previous Section.

\subsection{The Parisi function in the finite-size 2D XY model}

In order to compute the Parisi function $P(q,L)$ 
for the 2D XY model, one has to carefully 
define the overlap $q$.
Two different
overlaps between configurations (1) and (2) may be defined, namely
\begin{equation}
q^{12}  =  \frac{1}{L^2} \sum_{i=1}^{L^2} \cos [ \theta_i^1 - \theta_i^2 ]
\label{lab}
\end{equation}
or
\begin{equation} 
\tilde{q}^{12}  = \frac{1}{L^2} \sum_{i=1}^{L^2} \cos [ \theta_i^1 -\psi^1 
- \theta_i^2 + \psi^2 ],
\label{nolab}
\end{equation}
where $\psi^a = N^{-1} \sum_i \theta_i^a$.
According to the prescription of Ref.~\cite{silvio}, 
the $O(2)$ symmetry of the Hamiltonian has to be taken into account.
We have therefore considered the first definition. 
The Parisi function $P(q,L)$ can be
related to the probability distribution function of 
the magnetization $Q(m,L)$, where
\begin{equation}
m= \frac{1}{N} \sum_{i=1}^{N} \cos [ \theta_i - \psi ] \,.
\end{equation} 
The function $Q(m,L)$ gives   
the magnetization fluctuations in the direction defined by 
the mean orientation $\psi$, which contains the physics
of the critical fluctuations~\cite{peter}.
It has been analytically computed in 
Refs.~\cite{peter}.
The two first moments $\langle m \rangle$ and 
$\sigma = \sqrt{\langle m^2 \rangle}$ of the distribution
scale as~\cite{archambault} 
\begin{equation}
\langle m \rangle \sim L^{-\eta(T)/2},
\quad \sigma \sim T \langle m \rangle \,.
\end{equation} 
The temperature and size dependence of the distribution
enters only these two moments, and $Q(m,L)$ takes the scaling 
form~\cite{peter,archambault}
\begin{equation}
Q(m,L) = \sigma^{-1} \,{\cal F} \left( \frac{m -\langle m \rangle}{\sigma} 
\right),
\end{equation} 
where ${\cal F}(x)$ is a scaling function whose functional from
is well approximated by a generalized Gumbel 
distribution~\cite{peter}:
\begin{equation}
{\cal F}(x) = K(e^{y-e^{y}})^{\pi/2}, \quad y = b(x-s),
\end{equation}
where $K$, $b$ and $s$ are numerical constants ensuring the normalization
of the distribution.

\begin{figure}
\begin{center}
\psfig{file=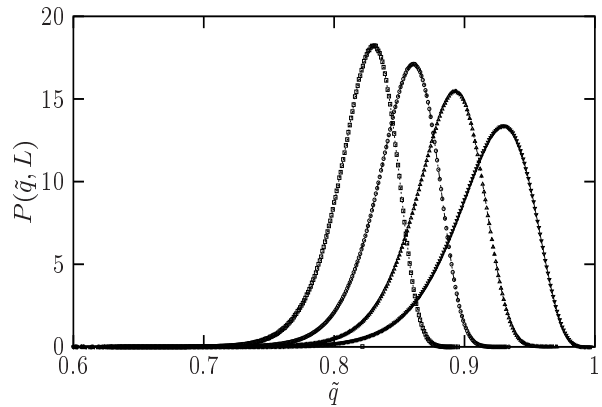,width=8.5cm,height=6.cm}
\caption{The Parisi function $P(\tilde{q},L)$ for the 2D XY model at $T=0.3$
and system sizes $L=4$, 8, 16, 32 from right to left.}
\label{pqnolab}
\end{center}
\end{figure}

From $Q(m,L)$, it is easy to obtain the critical fluctuations
of the overlap in the direction defined by the mean angle $\psi$, i.e.
using the definition (\ref{nolab}),
by the relation
\begin{equation}
P(\tilde{q},L) = \frac{1}{2\sqrt{\tilde{q}}}Q(\sqrt{\tilde{q}},L).
\end{equation}
This function is displayed in Fig.~\ref{pqnolab} for different system 
sizes.
Its scaling properties follow from those
of $Q(m,L)$~\cite{peter,archambault}
and are typical of the fluctuations of the order parameter
at a critical point~\cite{Binder}. The mean values scale with
system size as $\langle \tilde{q} \rangle \sim L^{-2\beta/\nu}$, 
as do the widths of the curves.
This result is expected from the hyperscaling relation between
critical exponents and is compatible
with a diverging susceptibility.
However it is not sufficient to 
give a finite amplitude for the probability at $\tilde{q}=0$. As
could be estimated from Fig.~\ref{pqnolab}, the probability function has an 
exponential tail for $q$ values below the mean, leading to much
larger fluctuations than for a Gaussian function, but the tail of
the distribution is still immeasurably small at $\tilde{q}=0$.

\begin{figure}
\begin{center}
\psfig{file=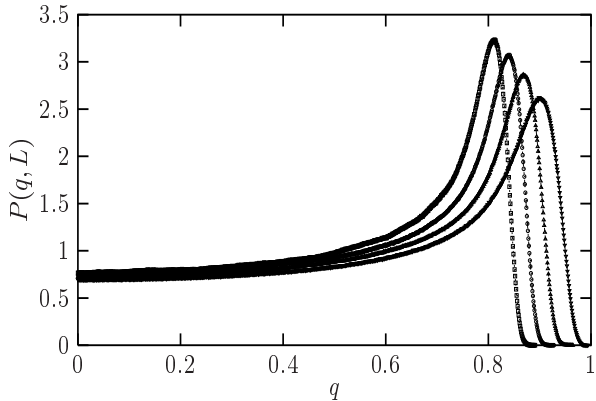,width=8.5cm,height=6.cm}
\caption{The Parisi function $P(q,L)$ for the 2D XY model at $T=0.3$
and system sizes $L=4$, 8, 16, 32 from right to left.}
\label{pqlab}
\end{center}
\end{figure}

When the definition (\ref{lab}) of the overlap is used,
a finite probability is generated at $q=0$ and 
a distribution reminiscent of the Parisi function for 
spin glasses is obtained~\cite{iniguez}. 
In this case the $O(2)$ symmetry is included
and $P(q,L)$ takes account of the fact that
the magnetization vector diffuses around the perimeter of a circle.
It is obtained from $Q(m,L)$ by the relation
\begin{equation}
P(q,L) = \frac{2}{\pi} 
\int_{\sqrt{q}}^1 \upd m \frac{Q(m,L)}{\sqrt{m^4-q^2}}.
\end{equation}
The result is shown in Fig.~\ref{pqlab} for different
system sizes. 
That the tail
is the product of this diffusive motion and is not the result
of critical fluctuations can be seen by considering the overlap 
between two rigid rotors of unit length diffusing on a circle, for which
$q=|\cos \theta |, \;\; q\ge 0$ and where $\theta$ is the 
angle between the two rod directions: this amounts
to computing $P(q,L)$ at $T=0$.
In this case, $P(q,L)$ is proportional 
to the density of states of $\cos \theta $ on
the circle and it follows easily that
\begin{equation}\label{QED}
P(q,L) = {2\over{\pi}}{1\over{\sqrt{1-q^2}}}.
\end{equation}
The intercept at $q=0$ in Fig.~\ref{pqlab} 
is quite close to the above value
of $2/\pi \simeq 0.64$, while the essential singularity at $q=1$ is removed
by the fluctuations in length of the magnetization vector. The critical
behaviour manifests itself again in the width and position of the maximum
probability, which scale with system size as in 
Fig.~\ref{pqnolab}. Apart from
this scaling of the peak, one would expect the same qualitative behaviour
for the low temperature phase of the 3D XY or 4D XY models, where symmetry
is broken in the thermodynamic limit.

While this explanation of the tail in $P(q,L)$ seems rather trivial it
is perhaps worth pointing out that the breaking of symmetry in an
XY model provides a very simple analog for the breaking of replica
symmetry in a mean field spin glass: the XY magnet contains an infinity 
of pure states $\alpha$ which are the points on the circle. On heating
and cooling in zero field the system samples different points on the
circle. It is true that these pure states are connected by a line of
constant free energy around the circle and so in principle the system can 
change pure states without jumping free-energy barriers. However, symmetry
breaking means that the diffusion constant around the circle goes to
zero in the thermodynamic limit and the system never leaves its chosen
pure state. Moving from one pure state to another across the circle
does involve jumping a free-energy barrier in complete analogy with the spin
glass. The overlap function between the pure ground states of the XY
ferromagnet is therefore given by the expression (\ref{QED}) above.
We note finally that distances between pure states of the 2D XY model 
on the circle do not verify the ultrametric inequality~\cite{parisi},
and hence are not organized in the hierarchical way they are in 
a mean-field spin glass.

\subsection{Finite-size statics vs finite-time dynamics}

We are now in position to test the relation (\ref{conjecture})
presented in the Introduction.
We compare the dynamic parametric plot Susceptibility/Correlation (data
of Fig.~\ref{param})
at two different temperatures, $T=0.7$ and $T=0.3$,
with the curves obtained through the relation (\ref{conjecture}).

\begin{figure}
\begin{center}
\psfig{file=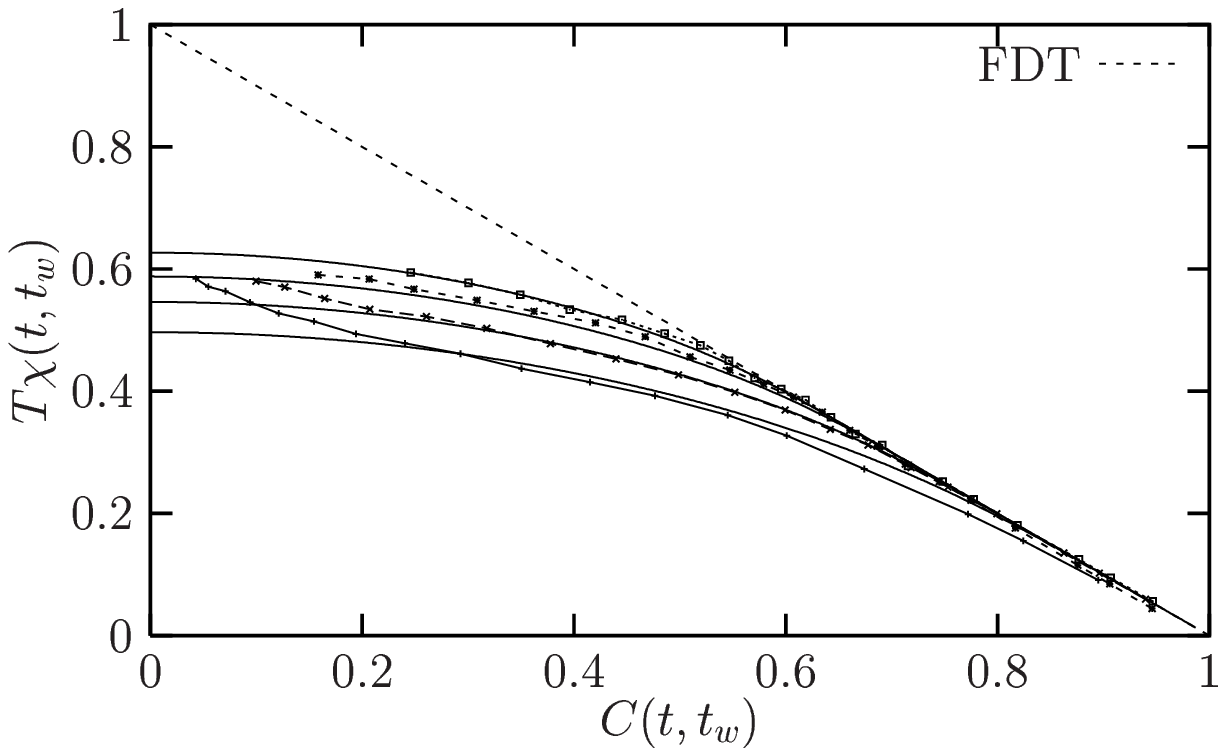,width=8.5cm,height=6.cm}
\caption{Comparison between the dynamical 
data and the curves built from the 
Parisi functions using the conjecture
(\ref{conjecture}) at $T=0.7$.
The linespoints are the dynamical data, while
the full lines are the static data.}
\label{parisi2}
\end{center}
\end{figure} 

At $T=0.7$, we find an excellent agreement between the dynamic and static
data (see Fig.~\ref{parisi2}).
This strongly supports our conjecture that the dynamic behaviour
can be described in terms of the static distribution
of a finite-size system of linear dimension such that $L=\xi(t_w)$.
There is a small systematic deviation between 
the two sets of curves on the left
hand side of the parametric plot.
This corresponds to the limit where the total simulation time
is a lot greater than $t_w$, which is most easily seen when $t_w$ is small.
In the regime where $C(t,t_w)$ is small and 
hence $\xi(t) \gg \xi(t_w)$, one can no longer
expect the relation (\ref{conjecture}) to be satisfied.

\begin{figure}
\begin{center}
\psfig{file=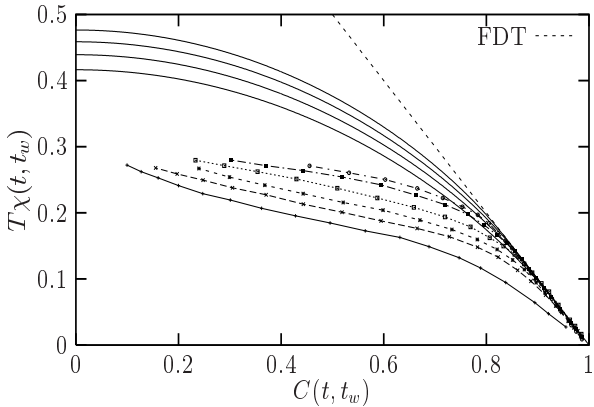,width=8.5cm,height=6.cm}
\caption{Comparison between the dynamical 
data and the curves built from the 
Parisi functions using the conjecture
(\ref{conjecture}) at $T=0.3$.
The linespoints are the dynamical data, while
the full lines are the static data.}
\label{parisi}
\end{center}
\end{figure}

At much lower temperature $T=0.3$, the two sets of curve coincide in the 
quasi-equilibrium regime $C>q_{EA}$ and the departure
point from the FDT is correctly predicted by the static data 
(see Fig.~\ref{parisi}).
As discussed above, this breaking point is given by
$C(t_w+t_w,t_w) \sim t_w^{-\eta(T)/2}$, which coincides
with the peak in the Parisi function $\langle q \rangle = L^{-2\beta/\nu}$,
with $L= t_w^{1/z}$.
Although the qualitative shape of the curves in the ageing regime
is the same for the two sets, there is a quantitative discrepancy.
This suggest that, while Eq.~(\ref{conjecture}) is a correct
starting point, the analytic $P(q,L)$ that we have calculated is not
in this case quantitatively accurate.

Indeed, in deriving this function we have explicitly neglected
vortices which play a crucial role in the transition from order to disorder
in the 2D XY model~\cite{koth}.
Since vortices are present during the coarsening process,
one should consider their contribution to the static 
fluctuations. 
We do not know precisely how this has to be done, for instance we do not know
what density and what vortex correlation to introduce.
However, it is clear that the presence of defects can dramatically 
alter the shape of the function $P(q,L)$.
As an illustration, we have calculated numerically $P(q,L)$
in the presence of a pinned defect.
We have slaved a single spin in the opposite direction
to its left neighbour and found that the tail at small $q$ is considerably
reduced.

\begin{figure}
\begin{center}
\psfig{file=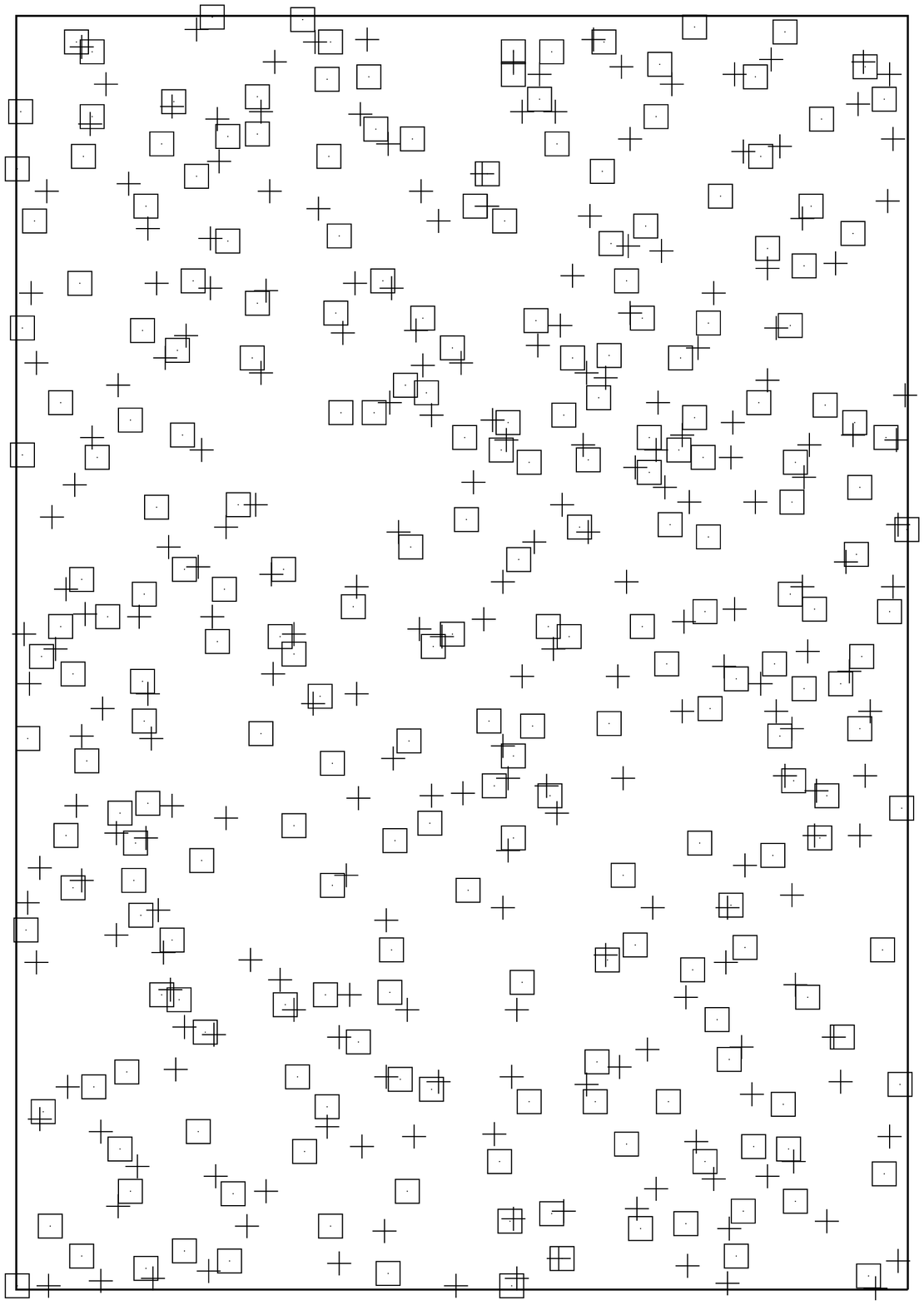,width=5.5cm,height=5.5cm} \\
\psfig{file=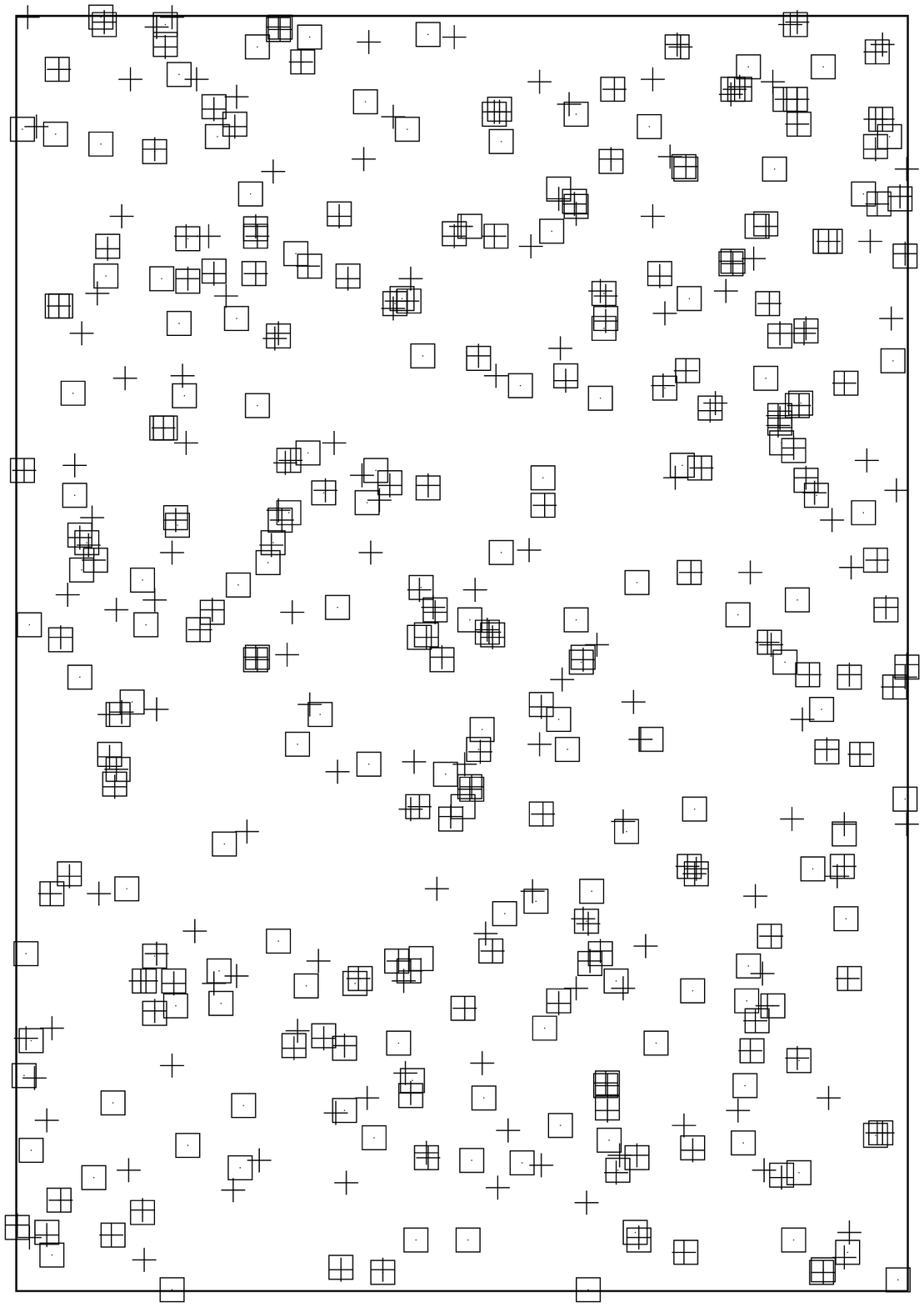,width=5.5cm,height=5.5cm}
\caption{Vortex configurations at time $t_w=10^3$ after a quench from a random
initial condition
to a temperature $T=0.3$ (above) and $T=0.7$ (below). 
Antivortices  are represented by squares, vortices
by plus symbols.}
\label{vortex}
\end{center}
\end{figure}

As a first step in investigating the effect of vortices, we show in 
Fig.~\ref{vortex} vortex configurations at the same waiting time, $t_w=10^3$,
for $T=0.3$ and $T=0.7$. 
While the number of vortices is of the same order (474 for $T=0.3$ and 602
for $T=0.7$), their correlations are clearly different.
At low temperature, the vortices of opposite charge are essentially
uncorrelated, while at higher temperature, the majority
of vortices are in closely bound pairs.
This is because the diffusion of vortices is more efficient at
higher temperature and so their natural tendency to form pairs
is largely satisfied.
The effect of a free vortex in the system is much more dramatic
than a vortex pair~\cite{koth}.
In Fig.~\ref{vortex}, there are far more free vortices 
at the lower temperature and hence one can expect that  
the static analysis in the absence of vortices
describes more accurately the situation at $T=0.7$ than at $T=0.3$.
This is consistent with our observations in Figs. \ref{parisi2}
and \ref{parisi}.

\section{Discussion}

We have studied in this work the nonequilibrium critical dynamics of the 
2D XY model and its relation with static properties.
We have investigated the ageing behaviour of the system by looking at
two-time dynamical functions and 
by considering quenches from random and ordered 
initial states to temperatures below the Kosterlitz-Thouless transition.
We have then shown that the finite-time violations of the FDT are 
related to the
finite-size equilibrium Parisi function. 
We discuss now some consequences of our results.

A first important feature is the scaling (\ref{autocorr})
of the spin-spin autocorrelation function.
Figure~\ref{CT03} shows that the behaviour
of $C(t,t_w)$ is, at first sight, very similar to the usual 
{\it additive} decomposition between a stationary and an ageing 
part~\cite{review_ageing}.  
We have seen that, at a critical point, this decomposition is
instead {\it multiplicative}.
Interestingly, the same scaling is observed in numerical
simulations of the 3D Edwards-Anderson model~\cite{review_simu,heiko}. 
It is indeed notoriously difficult to obtain the value of the Edwards-Anderson
parameter $q_{EA}$, the plateau in the correlation
function, in 3D Ising spin glasses~\cite{review_simu}.
This similarity means that the dynamics
of 3D spin glasses is influenced, at least at short time,
by the presence of a critical point. 

This multiplicative decomposition has also consequences 
in the scaling behaviour of the response function.
In the case of a quench in a ferromagnetic phase, the susceptibility
$\chi(t,t_w)$ can be decomposed into a `bulk' and an `interface'
contribution~\cite{BBK}.
The bulk contribution accounts for the equilibrium fluctuations
inside the domains, while the interface contribution accounts for the domain
wall response.
Since their density vanishes at long time, so does their response
function.
This is not true at the critical point, where there are no clearly
well defined `domains': on a length scale $\xi(t)$, the `bulk' is composed
of critical fluctuations, a self-similar structure
composed of domains within domains~\cite{HoHa,Ma,Binder}.
The direct consequence is that in a parametric Susceptibility/Correlation
plot, no master curve $\tilde{\chi}(C)$
(as predicted by the dynamical mean-field theory) can ever be 
reached.
The difference is irrelevant for the Ising spin chain at $T=0$, since 
there is no stationary part at all~\cite{golu,marco}.

We have conjectured a finite-time, finite-size generalization
of the relation (\ref{theorem}) in Eq.~(\ref{conjecture}) and 
shown that it works in the present model provided 
that the temperature is not very low.
We wish to mention that the conjecture (\ref{conjecture})
has been implicitly  used by Marinari {\it et al.} 
in Ref.~\cite{fdt_ea2}.
These authors have indeed shown that dynamic and static data 
for a 3D spin glass perfectly overlap, using the relation
(\ref{conjecture}) with $t_w=10^5$ and $L=16$.
Since $P(q,L)$ still depends on the size $L$ for $L=16$, and 
the $\chi$ {\it vs} $C$ plot still depends on $t_w$ at $t_w=10^5$
(because of the scaling (\ref{autocorr})),
the coincidence of the two sets of curves necessarily
results from Eq.~(\ref{conjecture}).
It is interesting to note that this correspondance
provides a direct measure of the dynamic correlation length $\xi(t)$.

As the temperature decreases, the agreement between the static 
and dynamic data is not as good because 
the density of free vortices increases.
This is consistent with the results for the Ising chain at 
$T=0$~\cite{marco}, where
Eq.~(\ref{theorem}) has been used to deduce from the dynamic
FDT violations a static `Parisi function'.
The latter is not linked  to the $T=0$ fluctuations 
of the spin glass order parameter which are trivial~\cite{regis}. 
Rather, the dynamic behaviour has been shown to be entirely controlled 
by the domain walls.
It would be interesting to see if a similar feature
occurs at low temperature in the 3D Edwards-Anderson model.

Since the ageing dynamics of the 2D XY model closely resembles 
that of more complex systems, it is important to understand why this is so.
We believe that the main ingredient for this behaviour
is the presence of the critical fluctuations.
After a quench, magnetic fluctuations of wave vectors between
$k \sim 1/a$ and $k\sim 1/L$ try to develop in a system
of size $L$ and lattice spacing $a$.
At time $t_w$, only the fluctuations with $k \gtrsim t_w^{-1/z}$ 
are equilibrated, while those with $k \lesssim t_w^{-1/z}$ are out
of equilibrium, the smallest $k$ being very far away from the 
equilibrium.
As a result, the decay of the autocorrelation function
is dictated by the growth of the correlation length which gives
the $\xi(t)/\xi(t_w)$-scaling of Eq.~(\ref{autocorr}).
Hence, this decay takes place in a single time scale.
However, the coexistence of multiple length scales, each 
having its own relaxation time, implies
FDT violations described by the smooth parametric plots of Fig.~\ref{param}.
The coexistence of a single time scale 
in the ageing regime together with a smooth $\chi$ vs $C$ plot arises 
naturally within this critical scenario.
This feature, observed in 3D and 4D spin glasses 
~\cite{fdt_ea2,fdt_ea,review_simu,heiko},
has no interpretation within the mean-field dynamical 
theory, where multiple effective temperatures are necessarily 
associated with multiple time scales~\cite{review_ageing,cuku,frme,BBK2}.

The relevance of multiple length scales in the ageing dynamics of spin glasses
has been emphasized by Bouchaud~\cite{trap}.
The example of the 2D XY model has shown that these length scales 
account well for the scaling properties of two-time dynamical functions
and FDT violations.
However, a necessary ingredient 
to explain more refined experimental results, such as the
`rejuvenation and memory' effects~\cite{memo}, lies in the role played by
the temperature~\cite{trap}.
In this picture, based on the behaviour of elastic objects pinned 
by disorder, the relaxation time $t(\ell)$
associated with a length scale $\ell$ scales as 
$t(\ell) \propto  \exp \left[ \Upsilon(T) \ell^\theta / T 
\right]$, in the notations of Ref.~\cite{trap}.
In the case of the 2D XY model the scaling is instead
$t(\ell) \sim \ell^z$,
independently of the temperature $T$.
This means that such a model will not be able to capture 
subtle temperature-cycling effects.
In this respect, it would be interesting to combine
the modified domain growth approach
recently proposed by Yoshino {\it et al.}~\cite{yoshino}
with nonequilibrium critical dynamics.

\section*{acknowledgments}     
We acknowledge useful discussions and correspondence 
with A. Barrat, J.-L. Barrat, A. J. Bray, J.-Y. Fortin,
S. Franz, J. Kurchan and J.-M. Luck.
MS is supported by the European Commission (contract ERBFMBICT983561).  
This work was supported by the P\^ole Scientifique de Mod\'elisation
Num\'erique at the \'Ecole Normale Sup\'erieure de Lyon.

\end{document}